\documentclass[aps,prl,twocolumn,superscriptaddress]{revtex4}

 \usepackage{bm}
 \usepackage{color}
 \usepackage{graphicx,psfrag} 
 \usepackage{amsmath,amssymb,mathrsfs}
 \usepackage{latexsym}

 \newcommand{\bea}{\begin{eqnarray}}
 \newcommand{\eea}{\end{eqnarray}}
 
 \newcommand{\braket}[2]{\left\langle #1 | #2 \right\rangle}
 \newcommand{\bra}[1]{\left\langle#1\right|}
 \newcommand{\ket}[1]{\left|#1\right\rangle}
 
 \newcommand{\of}[1]{\!\left(#1\right)}
 \newcommand{\sqof}[1]{\left[#1\right]}
 \newcommand{\cuof}[1]{\left\{#1\right\}}
 
 \newcommand{\abs}[1]{\left|#1\right|}
 
 \newcommand{\changed}[1]{#1}
 
 \newcommand{\bket}{\left\{z_1 \cdots z_{\num}\right\}}

 \newcommand{\expect}[1]{\left\langle#1\right\rangle}

 \newcommand{\Zcom}{\mathcal{Z}}
 \newcommand{\Wcom}{\mathcal{W}}
 
 \newcommand{\sith}{\mathrm{Co}}

\newcommand{\Kfunction}{K}
\newcommand{\Pfunction}{P}
\newcommand{\nfunction}{n}

\newcommand{\Wfunction}{W}

\newcommand{\Omegavec}{\bm{\Omega}}

\newcommand{\Svec}{\bm{S}}

\newcommand{\numelectrons}{M}
 \newcommand{\num}{\mathcal{N}}

\def\bd{\begin{displaymath}}
\def\ed{\end{displaymath}}
\def\be{\begin{equation}}
\def\ee{\end{equation}}
\def\bea{\begin{eqnarray}}
\def\eea{\end{eqnarray}}
\def\bi{\begin{itemize}}
\def\ei{\end{itemize}}
\def\bn{\begin{enumerate}}
\def\en{\end{enumerate}}

\def\ie{{\it i.e.},\ }

\newcommand{\zeroenergy}{zero-energy}
\newcommand{\PBCtranslation}{full lattice translation}

\begin{document}

\title{Spin Hamiltonian for which the Chiral Spin Liquid is the Exact Ground State}


\author{Darrell F. Schroeter}
\affiliation{Department of Physics, Reed College, Portland, OR 97202}
\author{Eliot Kapit}
\affiliation{Department of Physics, Cornell University, Ithaca,
NY, 14850}
\author{Ronny Thomale}
\author{Martin Greiter}
\affiliation{Institut f\"ur Theorie der Kondensierten Materie, 
Universit\"at Karlsruhe, D 76128 Karlsruhe}



\date{\today}

\begin{abstract}
  We construct a Hamiltonian that singles out the chiral
  spin liquid on a square lattice with periodic boundary conditions as
  the exact and, apart from the two-fold topological degeneracy, unique
  ground state.
\end{abstract}


\maketitle

{\it Introduction.} The fractionalization of quantum numbers, in which the excitations of a strongly-correlated system carry only a fraction of the quantum numbers of the constituents, is
currently of great interest in condensed matter physics and a
significant body of recent work has focused on finding solvable
theoretical models in which the phenomenon
occurs~\cite{motrunich-02prl277004, motrunich03prb115108,
  balents-02prb224412, misguich-02prl137202, moessner-01prl1881}.  In
addition to its intrinsic interest, the phenomenon of fractionalization may
well have a bearing on one of the most vexing problems in condensed
matter theory, should the long-standing suggestion of a link between
fractionalization and high-$T_{\mathrm{C}}$
superconductivity~\cite{anderson87s1197,kivelson-87prb8865} be
established; recently, it has been shown~\cite{ioffe-02n503} that the
topological degeneracy in these systems might be used to protect
quantum bits and be applicable to the emerging field of quantum
computing.

Fractional statistics, as a generalization of the idea of quantum
statistics based on Berry's phase~\cite{Wilczek90}, is a sensible idea
only in one or two dimensions, where one can define a winding number.
In 1D, the behavior is known to occur in spin-$1/2$
antiferromagnets~\cite{haldane91prl937, greiter06prl}, where exactly
solvable models exhibiting this behavior exist~\cite{haldane88prl635,
  shastry88prl639, haldane91prl1529, haldane-92prl2021,
  greiter-06prl}.  Fractionalization of statistics is also known to occur in 2D in
the presence of a magnetic field that violates the discrete
symmetries of parity $\of{\mathrm{P}}$ and time-reversal
$\of{\mathrm{T}}$; this situation is realized in the fractional
quantum Hall effect~\cite{laughlin83prl1395, stone92,
  halperin84prl1583, arovas-84prl722, camino-05prl246802,
  camino-05prb075342} (FQHE).  Very recently, the fractional
statistics of the quasiparticle excitations in the FQHE has been
observed experimentally~\cite{camino-05prl246802, camino-05prb075342}.
In contrast to the one-dimensional case, however, there has been no
definite evidence as to whether fractional statistics occurs in the absence of an external
field breaking these symmetries.

In this Letter, we present a spin Hamiltonian for which the chiral
spin liquid~\cite{kalmeyer-87prl2095, kalmeyer-89prb11879} (CSL) is
the exact ground state.  The CSL, one of the paradigmatic systems to introduce the concept of fractional statistics in 2D spin systems, is constructed to spontaneously
violate the symmetries $\mathrm{P}$ and $\mathrm{T}$; this violation is generally associated with fractional statistics.  The excitations of the liquid---spinons, which carry spin
$1/2$ but no charge, and holons, which carry charge but no spin---obey
fractional statistics.  In addition, the spinons exhibit quantum-number fractionalization and carry only half the spin of the excitations in conventional magnetically-ordered systems, which are spin-$1$. In many respects, the Hamiltonian we present is a generalization
of the Haldane-Shastry model~\cite{haldane88prl635, shastry88prl639}
(HSM) to 2D, and provides an exact spin model in which
fractional quantization can be studied.  
A spin Hamiltonian for a 2D system where the ground state is a chiral spin state, but not a liquid, has been discoverd by Wen {\it et al.}~\cite{wen-89prb11413}.  These authors additionaly argue for the  plausibility of a CSL ground state in a Heisenberg-like model including six-site interactions; the model presented in this Letter is precisely of that form.

The proof presented below shows that the model has an exact two-fold topologically degenerate ground state for \emph{any} number of lattice sites $N$.  This is in contrast to models, such as the Rokhsar-Kivelson dimer model~\cite{kivelson-87prb8865} (RKM), where the topological degeneracy is only realized in the thermodynamic limit~\cite{ioselevich-02prb174405}.  The exact topological degeneracy supports the view that the model will increase the accessibility of studying aspects of fractional statistics in 2D on an analytical and exact footing. At present, we have numerically verified the results in this Letter by exact diagonalization of the Hamiltonian on a $4\times4$ lattice. The numerical work confirms that the Hamiltonian has two zero-energy ground states and that these are the two topologically-degenerate CSL ground states introduced below.  A detailed discussion of the numerics will be presented elsewhere~\cite{manuscriptinpreparationTSG07}.  In the following, we brie!
 fly
review the CSL ground state, present the exact parent-Hamiltonian for the state, and show analytically that our positive
semi-definite Hamiltonian annihilates the CSL ground states.

\medskip {\it Ground state.} The CSL was originally conceived by
D.H.~Lee as a spin liquid constructed by condensing the bosonic spin
flip operators on a 2D lattice into a FQH liquid at Landau level
filling factor $\nu =1/2$.  The ground state wave function for a
circular droplet with open boundary conditions, on a square lattice
with lattice constant of length one, is given
by~\cite{kalmeyer-87prl2095, kalmeyer-89prb11879}
\begin{equation}
  \label{psiplane}
  \braket{z_1 \cdots z_{M}}{\psi}=
  \prod_{j<k}^M\,(z_j-z_k)^2\;\prod_{j=1}^M\,G(z_j)\,e^{-\frac{\pi}{2}|z_j|^2} \, ,
\end{equation}
where $\ket{\psi}$ is always referred to as the CSL state.
The $z$'s in the above expression are the complex positions of the
up-spins on the lattice: $z=x+iy$, with $x$ and $y$ integer.
$G(z)=(-1)^{(x+1)(y+1)}$ is a gauge factor, which ensures that
\eqref{psiplane} is a spin singlet.  Lattice sites not occupied by
$z$'s correspond to down-spins.

For our purposes, it is propitious to choose periodic boundary
conditions (PBCs) with equal periods $L_1=L_2=L$, $L$ even, and with
$N=L^2$ sites.  Following Haldane and Rezayi~\cite{haldane-85prb2529},
the wave function for the CSL then takes the form
\begin{eqnarray}
    \braket{z_1 \cdots z_{M}}{\psi}\! &\!=\!&\!
    \prod_{\nu =1}^2 \vartheta_1\of{\frac{\pi}{L}\sqof{\Zcom-Z_\nu}}
    \prod_{j<k}^M \vartheta_1\of{\frac{\pi}{L}\sqof{z_j-z_k}}^2 \nonumber\\
   &\cdot\!&\! \prod_{j=1}^M G(z_j)\,e^{\frac{\pi}{2}(z_j^2-|z_j|^2)},
  \label{eq:wavefunction}
\end{eqnarray}
where $M=N/2$ and $\vartheta_1(w)=-\vartheta_1(-w)
\equiv\vartheta_1\of{w|e^{-\pi}}$ is the odd Jacobi theta
function~\cite{AbramowitzStegun65}.  The zeros for the center-of-mass
coordinate $\Zcom=\sum_j z_j$ must lie in the principal region
$0\leq\text{Re}\of{Z_1}<L$, $0\leq\text{Im}\of{Z_1}<L$ and satisfy
$Z_1+Z_2=L+iL$; the freedom to choose $Z_1$ reflects the topological
degeneracy and yields two linearly independent ground states for the
CSL.  These states are spin singlets, are invariant under lattice
translations, and are strictly periodic with regard to the PBCs.

\medskip {\it Hamiltonian.} The Hamiltonian for which the CSL is the
exact ground state is defined in terms of vector operators
$\Omegavec_j$ as
\begin{eqnarray} 
  H =\sum_{\expect{i j}} \of{\Omegavec_i - \Omegavec_j}^{\dagger} \cdot
  \of{\Omegavec_i - \Omegavec_j} \, ,
  \label{eq:Hamiltonian}
\end{eqnarray} 
where the sum extends over all nearest neighbor pairs on the square
lattice.  The vector operators contain one-through-three-site
interactions and, in terms of two sets of coefficients $K_{i j k}$ and
$U_{ij}$, are defined as
\begin{widetext} 
  \begin{eqnarray} 
    \Omegavec_j = \sum_{i, k \neq j}' K_{ijk} \, \sqof{ \frac{1}{2
        \, i} \, \of{\Svec_j \times \Svec_k} + \frac{4}{5} \,
      \of{\Svec_j \cdot \Svec_k} \, \Svec_i - \frac{1}{5} \, \of{\Svec_k
        \cdot \Svec_i} \, \Svec_j - \frac{1}{5} \, \of{\Svec_i \cdot
        \Svec_j} \, \Svec_k} + \sum_{i \neq j} U_{ij} \, \Svec_i \, ,
	\label{eq:omegavecdef}
\end{eqnarray}
\end{widetext}
where the prime on the sum indicates $i\ne k$.
The coefficients $K_{i j k} = \Kfunction\of{z_k - z_j, z_i - z_j}$ in
the first term of~\eqref{eq:omegavecdef} are given by 
\begin{eqnarray}
  \Kfunction\of{x,y} = \lim_{R \rightarrow \infty} \sum_{0 \leq \abs{z_0
      - x} \leq R} \, \frac{1}{x - z_0} \, \frac{P\of{x - z_0, y}}{N/2 -
    1} \, ,
  \label{eq:Kdef}
\end{eqnarray} 
where the sum over all {\PBCtranslation}s $z_0 =\of{\ell + i\,
  m}L$ guarantees periodicity in the first argument of $K$. The
function $\Pfunction\of{x,y}$ is given by
\begin{eqnarray} 
  \Pfunction\of{x,y}\hspace{-2pt}=\hspace{-12pt}\sum_{0
    \leq \abs{z_0 - y}\leq R} \frac{\sith\of{\frac{\pi}{2L}
      \sqof{z_0 - y}}}{\sith\of{\frac{\pi}{2L}\sqof{x -
        \of{y - z_0}}}} \, \frac{e^{-\frac{\pi}{L^2}\abs{z_0
        - y}^2}}{\nfunction\of{y}},
  \label{eq:Pfunction}
\end{eqnarray} 
where $\sith\of{x} = \cos x + \cosh x$ and where $\nfunction\of{y}$ is
a normalization factor: \begin{eqnarray} \nfunction\of{y} = \lim_{R
    \rightarrow \infty} \sum_{0 \leq \abs{z_0 - y} \leq R} \,
  e^{-\frac{\pi}{L^2} \, \abs{z_0 - y}} \, ,
  \label{eq:nfunction}
\end{eqnarray} 
chosen such that $P\of{0,y} = 1$.  The sums in~\eqref{eq:Pfunction}
and~\eqref{eq:nfunction} enforce the periodicity of $\Kfunction$ in
its second argument.

The coefficients in the second term of~\eqref{eq:omegavecdef} are
given by $U_{ij} = \frac{\pi}{L} \, U\of{\frac{\pi}{L} \sqof{z_j -
    z_i}}$, where
\begin{multline}
  \label{eq:Udef}
\frac{\pi}{L} \, U\of{\frac{\pi}{L} z} = \frac{\pi}{L}\,
  \Wfunction\of{\frac{\pi}{L}\, z}+ \frac{1}{N-2} \\[5pt] 
  \cdot\Biggl[\left.
      \frac{d}{dx} P\of{x,-z} \right|_0 + \lim_{R \rightarrow \infty} \,
    \sum_{0 < \abs{z_0} \leq R} \frac{P\of{z_0,-z}}{z_0}\Biggr].
\end{multline}
In this expression, the function $W\of{z}$ is the periodic extension
of $1/z$ to the torus: 
\bea \frac{\pi}{L} W\of{\frac{\pi}{L} \, z} =
\lim_{R \rightarrow \infty} \sum_{0 \leq \abs{z_0} \leq R} \,
\frac{1}{z - z_0} \, .  
\label{eq:Wfunction}
\eea 
The Hamiltonian \eqref{eq:Hamiltonian} is
constructed to be positive semi-definite.  Therefore, if
$\Omegavec_i-\Omegavec_j$ annihilates the CSL states
\eqref{eq:wavefunction}, these states will be {\zeroenergy} ground
states of \eqref{eq:Hamiltonian}.

\medskip {\it Proof.} In order to prove that the vector operator
$\Omegavec_i - \Omegavec_j$ annihilates the CSL ground state, we first
demonstrate that the related tensor operator $\omega_i - \omega_j$
annihilates it.  Here, $\omega_i$
  is a reducible tensor, \ie a
composition of tensor components of different ranks, that may be
decomposed into irreducible spherical first-rank (vector) and
third-rank tensors; the operator $\Omegavec_i$ is the vector component
of $\omega_i$.  
\changed{The operator $\omega_i = \omega_i^+ - \omega_i^-$, where $\omega_i^{\pm}$ are related by a $\pi$-rotation about the $x$-axis, will be discussed in detail below after constructing the portion of the proof that does not depend on its precise form; it is later defined as $\omega_i^+ = T_i + V_i$ with the two operators $T_i$ and $V_i$ given in~(\ref{eq:tjdef}--\ref{eq:vjdef}) below.} 
The Wigner-Eckart theorem, in conjunction with the
fact that the ground state defined in~\eqref{eq:wavefunction} is a
spin singlet, guarantees that if $\omega_i - \omega_j$ is a
destruction operator for the state, then each of its irreducible
tensor components are also destruction operators.  Therefore, given that the operator $\omega_i - \omega_j$ destroys the ground
state, it follows that the vector operator $\Omegavec_i - \Omegavec_j$
does as well.  



In order to show that the operator $\omega_i - \omega_j$ is a
destruction operator for the ground state, we first demonstrate the
following property:
\begin{eqnarray} 
  \frac{\bra{z_1 \cdots z_{\numelectrons}} \, \omega_j
    \ket{\psi}}{\braket{z_1 \cdots z_{\numelectrons}}{\psi}} =
  f\of{\mathcal{Z}} \, .
  \label{eq:omegaproperty}
\end{eqnarray} 
The fact that the function on the right-hand side
of~\eqref{eq:omegaproperty} is independent of the site-index $j$
ensures that the difference of any two operators $\omega_i - \omega_j$
is a destruction operator.  We consider only
nearest-neighbor pairs of operators in constructing the Hamiltonian
in~\eqref{eq:Hamiltonian}, as this is the simplest and most local
operator that is also translationally invariant.  Other  choices,
however, are possible.

The reducible tensor operators $\omega_j$ can be decomposed into two
operators as $\omega_j = \omega_j^+ - \omega_j^-$, where $\omega_j^+$
and $\omega_j^-$ are related to each other through a $\pi$-rotation
about the $x$-axis that maps $S_z$ and $S_y$ into $-S_z$ and $-S_y$. 
\changed{The operator $\omega_j^+$ will be further decomposed as $\omega_j^+ = T_j + V_j$ with the explicit forms for these operators given in~(\ref{eq:tjdef}--\ref{eq:vjdef}) below.} 
In order to prove~\eqref{eq:omegaproperty}, we will first demonstrate
that
\begin{eqnarray}
  \frac{\bra{z_1 \cdots z_{\numelectrons}}\, 
    \omega_j^+ \ket{\psi}}{\braket{z_1 \cdots z_{\numelectrons}}{\psi}} 
  = f\of{\mathcal{Z}}\, 
  \left\{\begin{array}{ll} 1 & z_j \in \cuof{z_1 \cdots z_{\numelectrons}} \\
      0 & \mathrm{otherwise,} \end{array} \right.  
  \label{eq:omegaupproperty}
\end{eqnarray} 
where $f\of{\mathcal{Z}}$ is an odd, periodic function of the
center-of-mass coordinate $\Zcom$.  Using the relation between
$\omega_j^{\pm}$ and the invariance of the CSL ground state under such
a rotation, one can show, without specific knowledge of the function
$f\of{\Zcom}$, that 
\begin{eqnarray}
  \frac{\bra{z_1 \cdots z_{\numelectrons}} \, 
    \omega_j^- \ket{\psi}}{\braket{z_1 \cdots z_{\numelectrons}}{\psi}} 
  =f\of{\mathcal{\Wcom}} \, 
  \left\{\begin{array}{ll} 0 & z_j \in \cuof{z_1 \cdots z_{\numelectrons}} \\
      1 & \mathrm{otherwise.} \end{array} 
  \right.  
\end{eqnarray} 
In the above expression, $\Wcom = \sum w_i$ is the center of mass of
the down-spins on the lattice, such that $\cuof{w_i}$ is the
complement of $\cuof{z_i}$.  It is straightforward to show, regardless
of the chosen origin, that the sum of the two center of mass terms is
a {\PBCtranslation}: $\Zcom + \Wcom = \of{\ell + i \, m} \, L$.
This means that $f\of{\Wcom} = - f\of{\Zcom}$ and, given the
definition of $\omega_j$ above,~\eqref{eq:omegaproperty} follows
from~\eqref{eq:omegaupproperty}.

Having developed the remainder of the argument, it remains only to demonstrate~(\ref{eq:omegaupproperty}) for the operator $\omega_j^+$ to prove that~(\ref{eq:Hamiltonian}) is the exact parent Hamiltonian for the CSL; this last step is the heart of the proof.
The operator $\omega_j^+$ is defined in terms of off-diagonal and
diagonal contributions as $\omega_j^+ = T_j + V_j$ where
\begin{eqnarray}
  T_j & = & \frac{1}{2} 
  \sum_{i, k \neq j} K_{i j k} \, S_j^+ \, S_k^- \, \of{\frac{1}{2} + S_i^z} \label{eq:tjdef} \\
  V_j & = & \sum_{i \neq j}' U_{ij} \, \of{\frac{1}{2} + S_i^z}
  \, \of{\frac{1}{2} + S_j^z}, 
  \label{eq:vjdef}
\end{eqnarray} 
with the coefficients defined in terms of the functions in
(\ref{eq:Kdef}--\ref{eq:Udef}) above.  Considering first the
off-diagonal term, its action on the CSL ground state may be expressed as
\begin{widetext}
\begin{eqnarray}
	\bra{z_1 \cdots z_{\numelectrons}} T_j \ket{\psi} = \frac{1}{2} \sum_{i,k\neq j} K_{i j k} \bra{z_1 \cdots z_{\numelectrons}} S_j^+ \, S_k^- \, \of{\frac{1}{2} + S_i^z} \, \ket{\psi} \, .
\end{eqnarray}
This is clearly zero if $z_j \notin \bket$ giving half of the equality in~(\ref{eq:omegaupproperty}). Otherwise, acting onto the bra with the operator removes the site $z_j$ and replaces it with the site $z_k$.  In addition, the matrix element vanishes if $z_i \notin \bket$.  Dividing by the wave function yields
\bea
	\frac{\bra{z_1 \cdots z_{\numelectrons}} T_j \ket{\psi}}{\braket{z_1 \cdots z_{\numelectrons}}{\psi}} = \frac{1}{2} \sum_{i, k \neq j} K_{i j k} \frac{\braket{z_1 \cdots z_i \cdots z_k \cdots z_{\numelectrons}}{\psi}}{\braket{z_1 \cdots z_i \cdots z_j \cdots z_M}{\psi}} \, .
\eea
Using the definition in~\eqref{eq:Kdef} above, this may be written as
  \begin{eqnarray} \frac{\bra{z_1 \cdots z_{\numelectrons}} T_j \,
    \ket{\psi}}{\braket{z_1 \cdots z_{\numelectrons}}{\psi}} =
  \frac{1}{N - 2} \sum_{i \neq j} \, \sum_{z \neq 0} \lim_{R
    \rightarrow \infty} \sum_{0 \leq \abs{z_0 - z} < R} \,
  \frac{P\of{z - z_0, z_i - z_j}}{z - z_0} \, \frac{\braket{z_1 \cdots
      z_i \cdots z_j + z \cdots z_{\numelectrons}}{\psi}}{\braket{z_1
      \cdots z_i \cdots z_j \cdots z_{\numelectrons}}{\psi}}, 
      \label{eq:presumZ}
\end{eqnarray}
where the sum
over $k$ has been replaced by a sum over $z = z_k - z_j$.  As the wave function is periodic in all of its $M$
coordinates, the sums on $z_0$ and $z$ may be replaced with a sum on
$x = z - z_0$ that runs over the entire complex plane.  Additionally, the ratio of wave function coefficients appearing in (\ref{eq:presumZ}) has
the form
\begin{eqnarray}
  \frac{\braket{z_1 \cdots z_i \cdots z_j + x \cdots
      z_{\numelectrons}}{\psi}}{\braket{z_1 \cdots z_i \cdots z_j \cdots
      z_{\numelectrons}}{\psi}} & = & - G\of{x}  F\of{x} 
  e^{-\frac{\pi}{2} \, \abs{x}^2} \\
%
	F\of{x} & = & e^{\pi \, \of{z_j - z_j^*} \, x} \, e^{-\pi \, x^2 / 2} \prod_{i = 1}^2 \frac{\vartheta_1\of{\frac{\pi}{L} \, \sqof{\Zcom - Z_i + x}}}{\vartheta_1\of{\frac{\pi}{L} \, \sqof{\Zcom - Z_i}}} \prod_{k \neq i}^{M} \, \frac{\vartheta_1^2\of{\frac{\pi}{L}  \sqof{z_j - z_k + x}}}{\vartheta_1^2\of{\frac{\pi}{L}  \sqof{z_j - z_k}}} \, , 
\eea
where $F\of{x}$ is an analytic function of $x$. Being careful to pick up the points excluded by the sum in (\ref{eq:presumZ}) and using the above definition of $F\of{x}$, 
 the action of $T_j$ on the CSL ground state may be written as
  \begin{eqnarray} \frac{\bra{z_1 \cdots z_{\numelectrons}} T_j \,
    \ket{\psi}}{\braket{z_1 \cdots z_{\numelectrons}}{\psi}} = -
  \frac{1}{N - 2} \, \sum_{i \neq j} \sqof{ \sum_{x \neq 0} \,
    \frac{P\of{x, z_i - z_j}}{x} \, F\of{x} \, G\of{x} \,
    e^{-\frac{\pi}{2} \, \abs{x}^2} + \sum_{0 < z_0} \,     \label{eq:presumrule}
    \frac{P\of{z_0, z_i - z_j}}{z_0}} \, .  \end{eqnarray}
 This is the major step in the proof since the  first term may be evaluated with the singlet
sum-rule~\cite{laughlin89ap163}. This sum meets the requirements for
convergence~\cite{schroeter04ap155} that were not satisfied in the original
work, due to the exponential fall-off of $P\of{x,y}$ with increasing
$x$.  This gives 
\bea
	\sum_{x \neq 0} \,
    \frac{P\of{x, z_i - z_j}}{x} \, F\of{x} \, G\of{x} \,
    e^{-\frac{\pi}{2} \, \abs{x}^2} = - \left. \frac{d}{d x} \sqof{\frac{P\of{x, z_i - z_j}}{x} \, F\of{x}} \right|_{x = 0} \, . 
\eea
\end{widetext}
Combining this with the second term in Equation~\ref{eq:presumrule} gives
\begin{eqnarray} \frac{\bra{z_1 \cdots z_{\numelectrons}} T_j \,
    \ket{\psi}}{\braket{z_1 \cdots z_{\numelectrons}}{\psi}} =
  f\of{\mathcal{Z}} - \sum_{i \neq j} \frac{\pi}{L} \,
  U\of{\frac{\pi}{L} \, \sqof{z_j - z_i}},
  \label{eq:Taction}
\end{eqnarray} 
where the function of the COM coordinate $f\of{\Zcom}$, which first appears in (\ref{eq:omegaproperty}), may now be written down explicitly: 
\begin{eqnarray} f\of{\Zcom} = - \sum_{i=1}^2
  \frac{\pi}{2 \, L} \, W\of{\frac{\pi}{L} \, \sqof{\Zcom - Z_i}}  \, . 
\end{eqnarray} 
The $U$-function appearing in (\ref{eq:Taction}) is the one introduced  in (\ref{eq:Udef}) when defining the Hamiltonian.  The equality here is a result of the fact that  the $\Wfunction$-function, introduced in (\ref{eq:Wfunction}), is related to the logarithmic derivative of the odd Jacobi theta  functions:
\bea
	W\of{z} = \frac{d}{dz} \ln \vartheta_1\of{z} + \frac{z - z^*}{\pi} \, .
\eea

The operator $V_j$ introduced in (\ref{eq:vjdef}), which only generates diagonal terms, is chosen to exactly cancel the second term
in~\eqref{eq:Taction}.  This  proves the identity
in~\eqref{eq:omegaupproperty} and from here, the arguments at the
beginning of the section may be traced backwards to show that
$\Omegavec_i - \Omegavec_j$ annihilates the CSL state
\eqref{eq:wavefunction}, and hence that the CSL is an exact ground state
of \eqref{eq:Hamiltonian}.  The fact that the topological degeneracy is exact for any number of lattice sites $N$ in this model is due to the fact that $\omega_i - \omega_j$ destroys the state regardless of the choice of the location of the center-of-mass zeroes $Z_i$.

\medskip {\it Conclusion.}  We have constructed a Hamiltonian that
singles out the chiral spin liquid state as the exact and, apart from
the topological two-fold degeneracy for PBCs, unique zero-energy
ground state.  The proof has been numerically verified on a $4 \times 4$ lattice. In analogy to the HSM in one dimension, this model
provides a framework to study spinon excitations and their
interactions in a two-dimensional spin liquid.  For example, one may
investigate whether the spinons in this model are similarly free in the
sense that they only interact through their fractional statistics,
and, if so, whether the many spinon states can be classified in
similar terms~\cite{greiter-06prl}.  In any event, we have promoted the
CSL from an intriguing trial wave function to the exact ground state of a spin-Hamiltonian, and hence accomplished something analogous to the promotion of
Gutzwiller's wave function~\cite{gutzwiller63prl159} to an exact
solution by Haldane and Shastry~\cite{haldane88prl635,
  shastry88prl639}.
%

\begin{acknowledgments}
  DS acknowledges support from the Research Corporation under grant
  CC6682; RT was supported by the Studienstiftung
  des deutschen Volkes.  We would like to thank J.S. Franklin and R.B.
  Laughlin for many useful discussions.
\end{acknowledgments}


\end{document}